\documentclass[aps,showpacs,preprintnumbers,amsmath,amssymb]{revtex4}

\usepackage{float}
\usepackage{graphics,epsfig}
\usepackage{graphicx}
\usepackage{dcolumn}
\usepackage{bm}

\begin{document}
\baselineskip=0.8 cm
\title{{\bf Quasinormal modes in the background of charged Kaluza-Klein black hole
with squashed horizons}}
\author{Xi He}\affiliation{Department of Physics, Fudan
University, 200433 Shanghai}

\author{Songbai Chen}
\email{csb3752@163.com} \affiliation{Department of Physics, Fudan
University, 200433 Shanghai,
 \\ Institute of Physics and  Department of Physics,
Hunan Normal University,  Changsha, 410081 Hunan}

\author{Bin Wang}
\email{wangb@fudan.edu.cn} \affiliation{Department of Physics,
Fudan University, 200433 Shanghai}

\author{Rong-Gen Cai}
\email{cairg@itp.ac.cn} \affiliation{Institute of Theoretical
Physics, Chinese Academy of Sciences, P.O.Box 2735, 100080
Beijing}

\author{Chi-Yong Lin}
\email{lcyong@mail.ndhu.edu.tw} \affiliation{Department of
Physics, National Dong Hwa University, Shoufeng, 974 Hualien}

\vspace*{0.2cm}
\begin{abstract}
\baselineskip=0.6 cm
\begin{center}
{\bf Abstract}
\end{center}
We study the scalar perturbation in the background of the charged
Kaluza-Klein black holes with squashed horizons. We find that the
position of infinite discontinuities of the heat capacities can be
reflected in quasinormal spectrum. This shows the possible
non-trivial relation between the thermodynamical and dynamical
properties of black holes.
\end{abstract}

\pacs{ 04.70.Dy, 95.30.Sf, 97.60.Lf } \maketitle
\newpage

Quasinormal mode (QNM) of black holes has been an intriguing
subject of discussions for the last three decades \cite{1, 2, 3}.
It is believed that QNM is a characteristic sound of black holes
which could lead to the direct identification of the black hole
existence through gravitational wave observations to be realized
in the near future\cite{1, 2}. In addition to its potential
astrophysical interest, theoretically QNM was believed as a tool
to learn more about black hole and was even argued as a testing
ground for fundamental physics. It was found that the QNMs of
anti-de Sitter(AdS) black holes have direct interpretation in
terms of the dual conformal field theory(CFT)\cite{3, 4, 5, 6, 7,
8, 9}. This could serve as a support of the AdS/CFT
correspondence. Attempts of using QNMs to investigate the dS/CFT
correspondence have also been proposed\cite{10}. Recently it was
argued that QNMs might reflect the possible connection between the
classical vibrations of a black hole spacetime and various quantum
aspects by relating the real part of the QNM frequencies to the
Barbero-Immirzi(BI) parameter, which was introduced by hand in
order that loop quantum gravity reproduces correctly the black
hole entropy \cite{11,12}. But the direct connection has not been
found in AdS black hole background \cite{13}.

It is of great interest to investigate whether QNM can reflect
more physics of black holes. Recently some indications have been
found that black hole phase transitions can show up in the QNM
spectrum \cite{14,15,16}. This is interesting since it might be
the first phenomenon telling us the existence of the phase
transition in black hole physics. In order to examine whether the
QNM is an effective probe of phase transitions, we need to
investigate more general black hole configurations with more
general field perturbations. Jing et al \cite{17} computed the QNM
of Reissner-Nordstrom (RN) black hole and claimed that they found
the second order phase transition point predicted by Davies
\cite{18} where the heat capacity appears singular. However in
\cite{19}, it was argued that the result in \cite{17} might
probably be a numerical coincidence and the conjectured
correspondence between QNM and Davis's phase transition does not
straightforwardly generalize to Kerr or Schwarzschild AdS metrics.
Whereas calculations in \cite{19} cannot rule out the relation
between dynamical and thermodynamical properties of black holes
but suggest that such a relation is non-trivial.

In the study of black hole phase transition, there have been a lot
of debate on Davies' phase transition point. Around Davies' point,
the black holes' event horizons do not lose their regularities and
internal states of black holes do not change significantly. It was
considered more reasonable that the phase transition of black hole
occurs when a nonextremal black hole approaches its extremal
limit, since all second moments of non-equilibrium fluctuations
diverge there characterizing the second order phase transition
\cite{20}. Moreover extremal black holes are very different from
nonextreme holes. Extremal holes just have superradiation but
without Hawking radiation since its Hawking temperature vanishes.
The extremal black holes' geometric structures are also different
from their nonextremal counterparts, since the singularity will be
naked beyond the extremal limit. Davies' points for the divergence
of the heat capacity were later proved in fact as turning points
which are related to the changes of thermodynamical stability,
especially in the canonical ensemble \cite{21}. Whether the
thermodynamical stability can be reflected from the dynamical
stability such as the QNM spectra has been under heated
discussions recently (for a review see \cite{22}). For a black
brane solution, it was argued that the thermodynamical stability
is often related to dynamical stability\cite{23}. But whether this
correspondence is profound in all spacetimes is still not clear.

In this work we are going to study the QNM in the background of
charged Kaluza-Klein(KK) black hole with squashed horizons
\cite{24} and investigate whether the QNM spectra can reflect the
Davies' points of thermodynamical stability in this background. In
\cite{25}, the thermodynamic properties of the KK black hole are
discussed and compared to its undeformed five-dimensional RN black
hole counterpart. The Davies' thermodynamical stability point was
found where the heat capacity diverges when the horizon of the
black hole crosses the critical value. We will examine the scalar
perturbation in this background and see whether the thermodynamic
stability can be reflected in the QNM spectrum.

The five-dimensional charged KK black hole with squashed horizons
is described by\cite{24}
\begin{eqnarray}
ds^2=-f(r)dt^2+\frac{k^2(r)}{f(r)}dr^2+\frac{r^2}{4}k(r)d\Omega^2+\frac{r^2}{4}(d\psi+\cos\theta
d\phi)^2, \label{metric0}
\end{eqnarray}
where $d\Omega^2=d\theta^2+\sin^2\theta d\phi^2$ is the metric of
the unit sphere and
\begin{eqnarray}
f(r)=\frac{(r^2-r^2_+)(r^2-r^2_-)}{r^4},\;\;\;\;\;
k(r)=\frac{(r^2_{\infty}-r^2_+)(r^2_\infty-r^2_-)}{(r^2_{\infty}-r^2)^2}.
\end{eqnarray}
Here $0<\theta<\pi$, $0<\phi<2\pi$ and $0<\psi<4\pi$ are Euler
angles. The gauge potential is given by
\begin{eqnarray}
A=\pm\frac{\sqrt{3}}{2}\bigg(\frac{r_+r_-}{r^2}-\frac{r_+r_-}{r^2_{\infty}}\bigg)dt.
\end{eqnarray}
As in the RN black hole, the coordinate singularities $r=r_+$ and
$r=r_-$ correspond to the outer and inner horizons of the black
hole, respectively. $r_{\infty}$ is the spatial infinity. In the
parameter space $0 < r_-\leq r_+ < r_{\infty}$, $r$ is restricted
within the range $0<r<r_{\infty}$. The shape of black hole horizon
is deformed by the parameter $k(r_+)$.

In the metric (1), the intrinsic singularity is just the one at
$r=0$. This can be seen by introducing a new radial coordinate
$\rho$ as
\begin{eqnarray}
\rho=\rho_0\frac{r^2}{r^2_{\infty}-r^2},\label{p}
\end{eqnarray}
with
\begin{eqnarray}
\rho^2_0&=&\frac{k_0}{4}r^2_{\infty},\nonumber \\
k_0&=&k(r=0)=\frac{(r^2_{\infty}-r^2_+)(r^2_\infty-r^2_-)}{r^4_{\infty}}.
\end{eqnarray}
At the spatial infinity $r\rightarrow r_{\infty},
\rho\rightarrow\infty$. Thus in the new coordinate, $\rho$ varies
from $0$ to $\infty$ when $r$ changes from $0$ to $r_{\infty}$. The
metric (\ref{metric0}) can be rewritten as
\begin{eqnarray}
ds^2=-F(\rho)d\tau^2+\frac{K^2(\rho)}{F(\rho)}d\rho^2
+\rho^2K^2(\rho)d\Omega^2+\frac{r^2_{\infty}}{4K^2(\rho)}(d\psi+\cos\theta
d\phi)^2, \label{metric}
\end{eqnarray}
with
\begin{eqnarray}
F(\rho)=\bigg(1-\frac{\rho_+}{\rho}\bigg)\bigg(1-\frac{\rho_-}{\rho}\bigg),\;\;\;\;\;\;\;\;\;\;
K^2(\rho)=1+\frac{\rho_0}{\rho}.
\end{eqnarray}
Here we have defined the proper time $\tau=2\rho_0 t/r_{\infty}$ for
the observer at infinity. The mass and charge for this squashed KK
black hole are defined by
\begin{eqnarray}
M=\frac{3\pi
r_{\infty}}{4G}(\rho_++\rho_-),\;\;\;\;\;\;\;\;\;Q=\frac{\sqrt{3}\pi
r_{\infty}}{G}\sqrt{\rho_+\rho_-},
\end{eqnarray}
where $\rho_{\pm}=\rho_0r^2_{\pm}/(r^2_{\infty}-r^2_{\pm})$ are
the outer and inner horizons of the black hole in the new
coordinate.

The Hawking temperature and entropy of the black hole can be
expressed as \cite{25}
\begin{eqnarray}
&&T_H=\frac{\rho_+-\rho_-}
{4\pi\rho^2_+}\sqrt{\frac{\rho_+}{\rho_++\rho_0}}=\frac{r^2_+-r^2_-}{2\pi r^3_+}\frac{r^2_{\infty}}{r^2_{\infty}-r^2_-}\sqrt{\frac{r^2_{\infty}-r^2_+}{r^2_{\infty}-r^2_-}},\label{TH}\\
&&S=4\pi^2(\rho_+)^{\frac{3}{2}}(\rho_0+\rho_-)^{\frac{1}{2}}(\rho_0+\rho_-)=\frac{\pi^2
r^3_+}{2}\frac{r^2_{\infty}-r^2_-}{r^2_{\infty}-r^2_+}.
\end{eqnarray}

Several limits of this charged squashed KK black hole were
discussed in \cite{24,25}. When $r_{\infty}\rightarrow \infty$,
the squashing function $k$ in (2) tends to be unity so that the
metric (1) reduces to that of five-dimensional RN black hole. The
entropy and the Hawking temperature also reduce to those of five-
dimensional RN cases. When $r_-$ goes to zero, we have the metric
for the neutral black hole with squashed horizon. If we have
$r_-=r_+$, we have deformed five-dimensional extremal RN black
hole with one degenerate horizon whose Hawking temperature
vanishes. For the case $r_+, r_-\rightarrow r_{\infty}$, with
$\rho_{\pm}$ finite, it is convenient to see from (6) that because
$\rho_0\rightarrow 0, K^2(\rho)\rightarrow 1$, the metric
describes the four-dimensional RN black hole with a twisted $S^1$
bundle, where the size of the $S^1$ fiber takes the constant value
$\sqrt{\rho_+\rho_-}=r_{\infty}/2$\cite{24}. Its temperature
reduces to that of a four-dimensional black hole in this limit.

The heat capacity of the black hole for the fixed $Q$ is given
by\cite{25}
\begin{eqnarray}
C_{Q}=T \bigg({\partial S\over \partial T}\bigg)_Q={\pi^2 r^3_+
\over 2} {r^2_\infty- r^2_- \over r^2_\infty -r^2_+}{(r^2_+
-r^2_-)(3r^4_\infty -r^2_+ r^2_- -r^2_\infty (r^2_+ +r^2_-))\over
r^4_\infty (5 r^2_- -r^2_+)-r^2_- (2r^2_\infty -r^2_+)(3 r^2_+
+r^2_-)}.
\end{eqnarray}
Since $3r^4_\infty -r^2_+ r^2_- -r^2_\infty (r^2_+ +r^2_-)>0$, the
sign of the heat capacity $C_Q$ is determined by the term
$r^4_\infty (5 r^2_- -r^2_+)-r^2_- (2r^2_\infty -r^2_+)(3 r^2_+
+r^2_-)$ in the denominator. It was found that the Davies' point
exists at $r_{crit}$ \cite{25} and one has $C_Q>0$ for
$r_+<r_{crit}$ while $C_Q<0$ for $r_+>r_{crit}$ and $C_Q$ diverges
when $r_+$ crosses this critical value $r_{crit}$. In the new
coordinate, the Davies' point of the divergence of  heat capacity
$C_Q$ can be obtained from
\begin{eqnarray}
\rho_0(\rho_+-5\rho_-)+
(\rho_+-3\rho_-)(\rho_++\rho_-)=0.\label{c1}
\end{eqnarray}

If we take $b=\rho_+-\rho_-$, the above equation (\ref{c1}) can be
rewritten as
\begin{eqnarray}
b=\frac{(\rho_++\rho_-)(\rho_++\rho_-+2\rho_0)}{2(\rho_++\rho_-)+3\rho_0}.
\end{eqnarray}
In the limit $\rho_0\rightarrow0$, we have $b=(\rho_++\rho_-)/2$
and $\rho_+=3\rho_-$, which is the Davies' point of thermal
stability in the four-dimensional RN black hole. When
$\rho_0\rightarrow\infty$,
 i.e., $r_{\infty}\rightarrow\infty$, we obtain $b=2/3(\rho_++\rho_-)$ and $\rho_+=5\rho_-$,
which is consistent with the Davies' point in five-dimensional RN
black hole.

In the following we are going to investigate whether the Davies'
points on thermal stability can be reflected in QNM. We will
concentrate on the massless scalar perturbation around the charged
KK black hole with squashed horizons. The wave equation for the
massless scalar field $\Phi(\tau,\rho,\theta,\phi,\psi)$ in the
background (\ref{metric}) obeys
\begin{eqnarray}
\frac{1}{\sqrt{-g}}\partial_{\mu}(\sqrt{-g}g^{\mu\nu}\partial_{\nu})
\Phi(\tau,\rho,\theta,\phi,\psi)=0.\label{WE}
\end{eqnarray}
Taking the ansatz $\Phi(\tau,\rho,\theta,\phi,\psi)=e^{-i\omega
\tau}R(\rho)e^{i m\phi+i\lambda \psi}S(\theta)$, where $S(\theta)$
is the so-called spheroidal harmonics, we can obtain the equation
\begin{eqnarray}
\frac{1}{\sin{\theta}}\frac{d}{d\theta}\bigg[\sin{\theta}
\frac{d}{d\theta}\bigg]S(\theta)
-\bigg[\frac{(m-\lambda\cos{\theta})^2}{\sin^2{\theta}}-E_{lm\lambda}\bigg]S(\theta)=0,\label{angd}
\end{eqnarray}
for the angular part. The eigenvalue of this angular equation
(\ref{angd}) is $E_{lm\lambda}=L(L+1)-\lambda^2$. The radial
equation reads
\begin{eqnarray}
\frac{F(\rho)}{\rho^2K^2(\rho)}\frac{d}{d\rho}\bigg[\rho^2F(\rho)\frac{d
R(\rho)}{d\rho}\bigg] +[\omega^2-V(\rho)]R(\rho) =0,\label{radial}
\end{eqnarray}
with
\begin{eqnarray}
V(\rho)
=F(\rho)\bigg(\frac{L(L+1)-\lambda^2}{\rho^2K^2(\rho)}+\frac{4\lambda^2K^2(\rho)}{r^2_{\infty}}\bigg).
\end{eqnarray}
The second term in the effective potential came from the fifth
dimension of the spacetime. It plays a role of the mass of the field
in the radial equation. In general, due to the presence of this
term, it is difficult to calculate the QNM through continued
fraction method.

Boundary conditions on the wave function $R(\rho)$ at the outer
horizon and the spatial infinity can be expressed as
\begin{eqnarray}
R(\rho)\sim \left\{
  \begin{array}{ll}
    {(\rho-\rho_+)^{\frac{i
\rho_+^{3/2} \sqrt{\rho_0+\rho_+} w}{\rho_--\rho_+}}},
&{\hbox{$\rho$ $\rightarrow$ $\rho_+$};}
\\ \\
    {\rho^{\frac{i (\rho_0+2 (\rho_-+\rho_+)) w^2}{2\chi }-\frac{i (2
\rho_0+\rho_-+\rho_+) \lambda
   ^2}{2(\rho_0+\rho_-) (\rho_0+\rho_+) \chi }-1}
e^{i\chi\rho}},~~~~ & \hbox{$\rho\rightarrow\infty$.}\label{bd1}
  \end{array}
\right.
\end{eqnarray}

A solution of equation (\ref{radial}) that satisfies the above
boundary condition can be written as
\begin{eqnarray}
R(\rho)&=&e^{i (\rho-\rho_-) \chi}
   (\rho-\rho_-)^{{\frac{i \rho_+^{3/2} \sqrt{\rho_0+\rho_-}
   \omega}{\rho_+-\rho_-}}+{\frac{i [\rho_0+2
(\rho_-+\rho_+)]
   \omega^2}{2\chi }-\frac{i (2 \rho_0+\rho_-+\rho_+) \lambda
   ^2}{2(\rho_0+\rho_-) (\rho_0+\rho_+) \chi }}} \\ \nonumber &&(\rho-\rho_+)^{\frac{i \rho_+^{3/2}
   \sqrt{\rho_0+\rho_+} \omega}{\rho_--\rho_+}}\sum_{m=0}^{\infty}a_m\bigg({\rho-\rho_+ \over \rho-\rho_-}\bigg)^m,
\label{so}
\end{eqnarray}
where $\chi^2=\omega^2-\frac{\lambda ^2}{(\rho_0+\rho_-)
(\rho_0+\rho_+)}$. The sequence of expansion coefficient ${a_m:
~m=1,~2,~3...}$ is determined by the recurrence relation starting
from $a_0=1$
 \begin{eqnarray}
&&\alpha_0 a_1 +\beta_0 a_0=0, \\ \nonumber && \alpha_m
a_{m+1}+\beta_m a_m + \gamma_m a_{m-1}=0, ~~~m=1,~2,...
 \end{eqnarray}
The recurrence coefficients $\alpha_m$, $\beta_m$, $\gamma_m$ are
given by
\begin{eqnarray}
&&\alpha_m=m^2+(C_0+1)m+C_0, \\ \nonumber
&&\beta_m=-2m^2+(C_1+2)m+C_3, \\ \nonumber
&&\gamma_m=m^2+(C_2-3)m+C_4-C_2+2,
\end{eqnarray}
where $C_m$ are
\begin{eqnarray}
C_0&=&\frac{2 i \sqrt{\rho_0+\rho_+} \omega \rho_+^{3/2}}{\rho_- -\rho_+}+1, \\
C_1&=& -\frac{4 i \sqrt{\rho_0+\rho_+} \omega
   \rho_+^{3/2}}{\rho_--\rho_+}+\frac{i \big[(\rho_0+\rho_-)
   \big(\rho_0^2+5 \rho_+ \rho_0+4 \rho_+^2\big) \omega^2+(-2 \rho_0+\rho_--3 \rho_+) \lambda ^2\big]}{
   (\rho_0+\rho_-) (\rho_0+\rho_+) \chi }-4,\\
C_2&=&3-\frac{i \big[\rho_0 +2 (\rho_-+\rho_+)\big] \omega^2}{\chi
   }+\frac{2
   i \rho_+^{3/2} \sqrt{\rho_0+\rho_+}
   \omega}{\rho_--\rho_+}+\frac{i \big(2 \rho_0+\rho_-+\rho_+\big) \lambda ^2}{(\rho_0+\rho_-) (\rho_0+\rho_+)
   \chi },\\ \nonumber
C_3&=&{1\over {2 (\rho_0+\rho_-)
   (\rho_--\rho_+) (\rho_0+\rho_+) \chi }}\bigg\{\lambda ^2 \bigg[2 (\rho_0+\rho_+) \chi
(\rho_--\rho_+)^2+  i (-2 \rho_0+\rho_--3 \rho_+)
   (\rho_--\rho_+) \nonumber\\
   && +2 \rho_+^{3/2} \sqrt{\rho_0+\rho_+} (2 \rho_0-\rho_-+3 \rho_+)
   \omega \bigg]-(\rho_0+\rho_-) (\rho_0+\rho_+) \bigg[\bigg(2 \rho_+^2 \big(3 \rho_0+4 \rho_+ \big) \chi -i
   \big(\rho_--\rho_+\big)\\ \nonumber&& \big(\rho_0+4 \rho_+\big)\bigg) \omega^2
   +2 \rho_+^{3/2} \sqrt{\rho_0+\rho_+} \left( (\rho_0+4
   \rho_+) \omega^2+2 i \chi \right) \omega+2 \left(L^2+L+1\right) (\rho_--\rho_+) \chi \bigg]\bigg\},\\
C_4&=&\frac{(\rho_0+\rho_-) \omega^2
   \rho_-^3}{(\rho_+-\rho_-)^2}+\frac{1}{4} \bigg\{\frac{2 i
   \rho_+^{3/2} \sqrt{\rho_0+\rho_+} \omega}{\rho_--\rho_+}-\frac{i
   \big[\rho_0+2
   (\rho_-+\rho_+)\big] \omega^2}{\chi }+\frac{i \big(2
   \rho_0+\rho_-+\rho_+\big) \lambda ^2}{(\rho_0+\rho_-) (\rho_0+\rho_+) \chi
   }+2\bigg\}^2,
\end{eqnarray}

If the boundary condition (\ref{bd1}) is satisfied and the series
in (\ref{so}) converge for the given $L$, the frequency $\omega$
is a root of the continued fraction equation
\begin{eqnarray}
\bigg[\beta_m-{\alpha_{m-1}\gamma_m\over
\beta_{m-1}-}{\alpha_{m-2}\gamma_{m-1}\over
\beta_{m-2}-}...{\alpha_0 \gamma_0 \over
\beta_0}\bigg]=\bigg[{\alpha_m \gamma_{m+1}\over
\beta_{m+1}-}{\alpha_{m+1}\gamma_{m+2}\over
\beta_{m+2}-}{\alpha_{m+2}\gamma_{m+3}\over
\beta_{m+3}-}....\bigg],\;\;\;(m=1,2,...).\label{cnf}
\end{eqnarray}
This means that we can calculate the QNM frequencies of the
charged KK black hole with squashed horizons by solving the above
continued fraction equation (\ref{cnf}). This equation is
impossible to be solved analytically. We can only rely on the
numerical calculation to obtain the QNM frequencies.

\begin{figure}
\includegraphics[width=13cm]{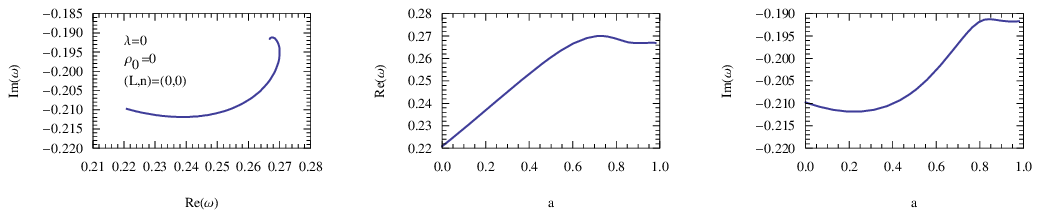}
\includegraphics[width=13cm]{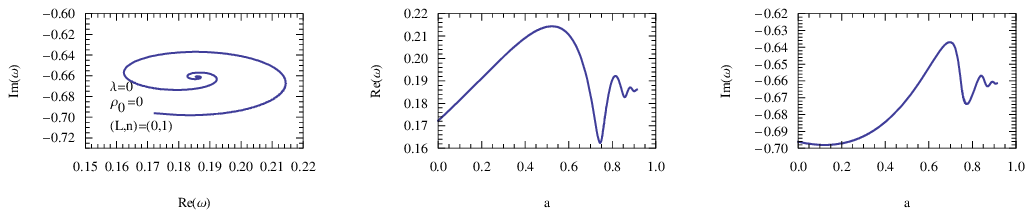}
\caption{\label{g000-010} The left two panels are trajectories in
the complex $\omega$ plane of scalar QNMs around squashed KK black
holes, for $n=0$, $n=1$, with $L=0$ and $\rho_0=0$. The others are
real parts ($Re(\omega)$) and imaginary parts $(Im(\omega)$)
versus parameter $a=1-b$.}
\end{figure}

\begin{figure}
\includegraphics[width=13cm]{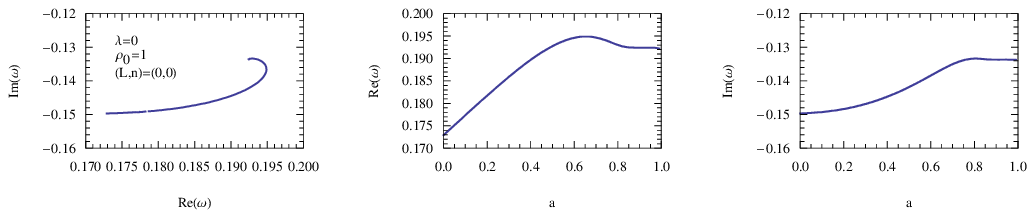}
\includegraphics[width=13cm]{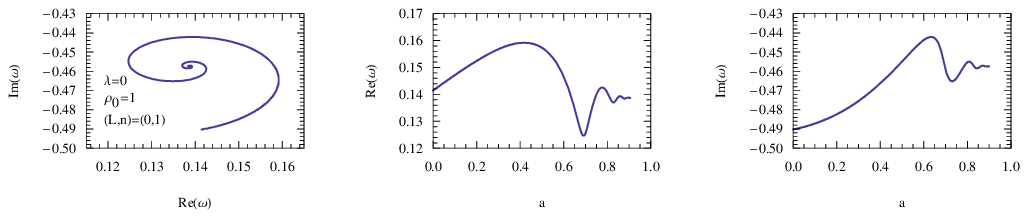}
\caption{\label{g001-011}  The left two panels are trajectories in
the complex $\omega$ plane of scalar QNMs around squashed KK black
holes, for $n=0$, $n=1$, with $L=0$ and $\rho_0=1$. The others are
real parts ($Re(\omega)$) and imaginary parts $(Im(\omega)$)
versus parameter $a=1-b$.}
\end{figure}

\begin{figure}
\includegraphics[width=13cm]{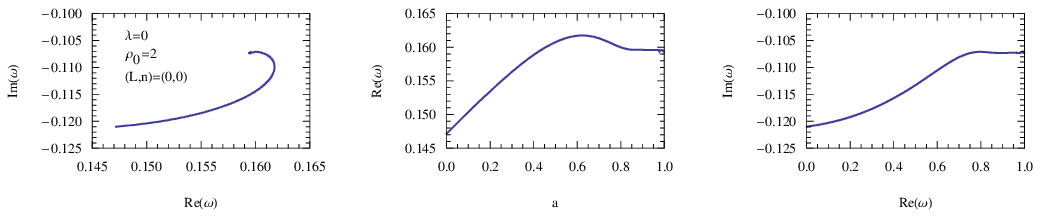}
\includegraphics[width=13cm]{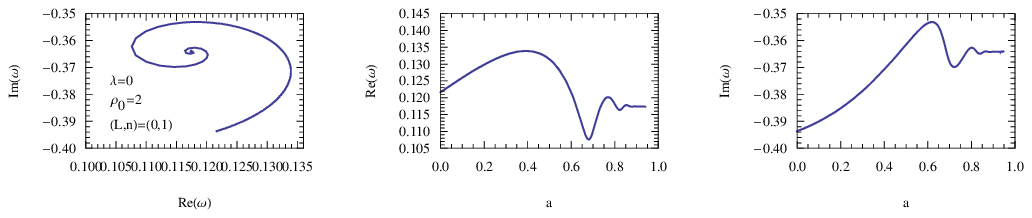}
\caption{\label{g002-012} The left two panels are trajectories in
the complex $\omega$ plane of scalar QNMs around squashed KK black
holes, for $n=0$, $n=1$, with $L=0$ and $\rho_0=2$. The others are
real parts ($Re(\omega)$) and imaginary parts $(Im(\omega)$)
versus parameter $a=1-b$.}
\end{figure}

Taking the limit $Q\rightarrow 0$, we have compared our numerical
results by using Leaver's method with that obtained in
\cite{Ishihara} by other methods. The comparison is shown in Table
\ref{compare}, which shows good agreement and in addition we see
that Leaver's method we adopted gives more precise result.

In Figs.~(\ref{g000-010})-(\ref{g002-012}) we display the QNM
frequencies of scalar perturbation around the squashed KK black
holes with charge for $L=0$, $\lambda=0$, with overtone numbers
$n=0,~1$ and $\rho_0=0,~1,~2$, respectively. We plot both the real
part and the imaginary part of QNM frequencies in functions of
$a=1-b$. Black hole horizons are related to $a$ by $\rho_+=1-a/2,
\rho_-=a/2$. We observe that over the critical overtone number
($n_c=1$ for $L=0$ in the scalar perturbation), both the real part
and the imaginary part of QNM frequencies will behave oscillatory
when $a$ crosses a critical value $a_{Q}$ and meanwhile the complex
$\omega$ plan will exhibit the spiral-like shape. For $\rho_0=0$,
which is the limiting case of the four-dimensional RN black hole
with a twisted $S^1$ bundle, we observed that the real part of the
QNM frequency arrives at its first maximum of its oscillation
approximately at the same position as the Davies' point $a_{D}=1/2$.
The difference between the critical point from the QNM $a_{Q}$ from
that of the Davies' point is very small, $|a_{Q}-a_{D}|=0.022$. For
$\rho_0$ not equaling to zero, the Davies' thermal stability point
$a_{D}$ can be calculated through (\ref{c1}) and from the behavior
of QNM frequencies we can read the critical value $a_{Q}$ when the
oscillation of the real and imaginary parts of frequencies start.
The results are shown in table \ref{error}. It is interesting to
find that critical points got from QNM agree very well to those of
Davies' points.

\begin{figure}
\includegraphics[width=13cm]{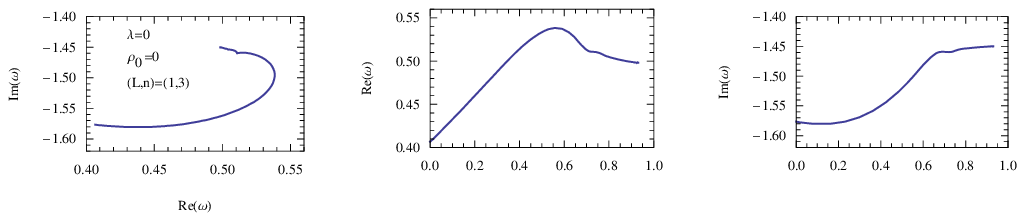}
\includegraphics[width=13cm]{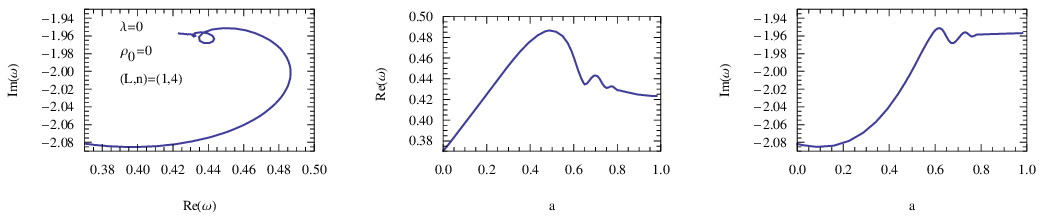}
\caption{\label{g130-040} The left two panels are trajectories in
the complex $\omega$ plane of scalar QNMs around squashed KK black
holes, for $n=3$, $n=4$, with $L=1$ and $\rho_0=0$. The others are
real parts ($Re(\omega)$) and imaginary parts $(Im(\omega)$)
versus parameter $a=1-b$.}
\end{figure}

\begin{figure}
\includegraphics[width=13cm]{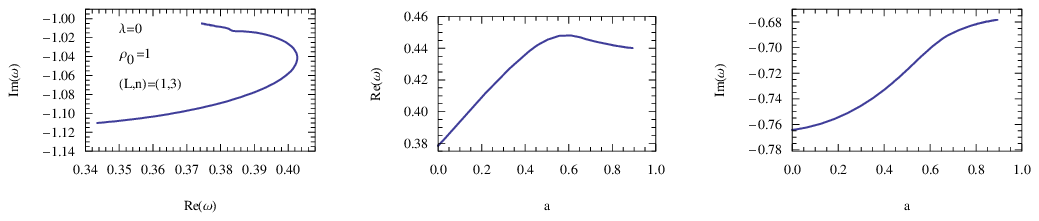}
\includegraphics[width=13cm]{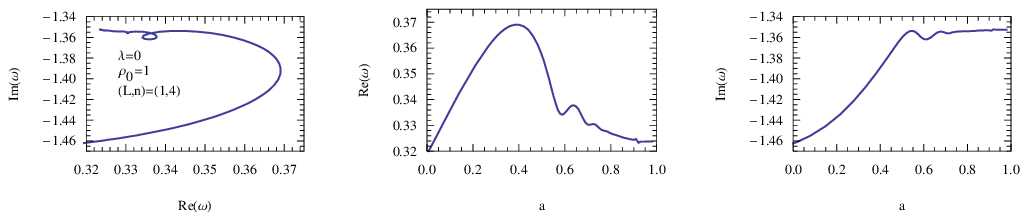}
\caption{\label{g131-141}The left two panels are trajectories in
the complex $\omega$ plane of scalar QNMs around squashed KK black
holes, for $n=3$, $n=4$, with $L=1$ and $\rho_0=1$. The others are
real parts ($Re(\omega)$) and imaginary parts $(Im(\omega)$)
versus parameter $a=1-b$.}
\end{figure}

\begin{figure}
\includegraphics[width=13cm]{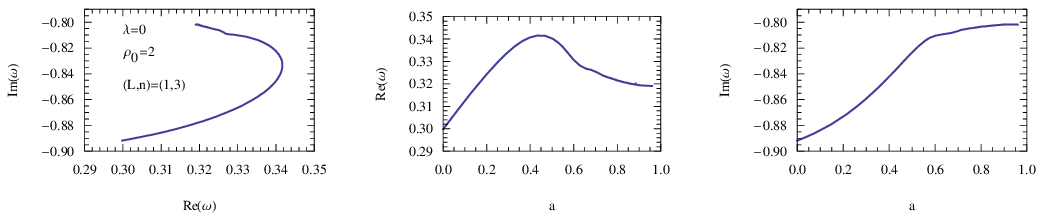}
\includegraphics[width=13cm]{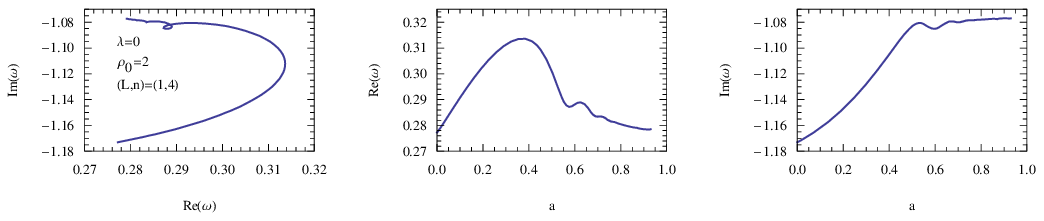}
\caption{\label{g132-142}The left two panels are trajectories in
the complex $\omega$ plane of scalar QNMs around squashed KK black
holes, for $n=3$, $n=4$, with $L=1$ and $\rho_0=2$. The others are
real parts ($Re(\omega)$) and imaginary parts $(Im(\omega)$)
versus parameter $a=1-b$.}
\end{figure}

In Figs.~(\ref{g130-040})-(\ref{g132-142}), we show results for
$L=1$ and $\lambda =0$, with overtone numbers $n=3,~4$ and
$\rho_0=0,~1,~2$, respectively. The Davies' points are the same for
each selected values of $\rho_0$ no matter the change of $L, n$. As
was observed in \cite{17}, with the increase of $L$, we will observe
the spiral-like shape of the complex $\omega$ plan and oscillatory
real and imaginary frequencies of QNM over higher critical overtone
number $n_c$, namely $n_c=4$ when $L=1$ for scalar perturbation. The
oscillations of the QNM frequencies start when $a$ over the critical
value $a_Q$, which is again very much in agreement with Davies'
critical point $a_D$. The results of their differences are shown in
table \ref{error}. Choosing very big value of $\rho_0$, we observed
that the critical point of $a$ from QNM will go towards $1/3$ which
is the Davies point for the five-dimensional RN black hole for
$\rho_0\rightarrow \infty$.

\begin{figure}
\includegraphics[width=13cm]{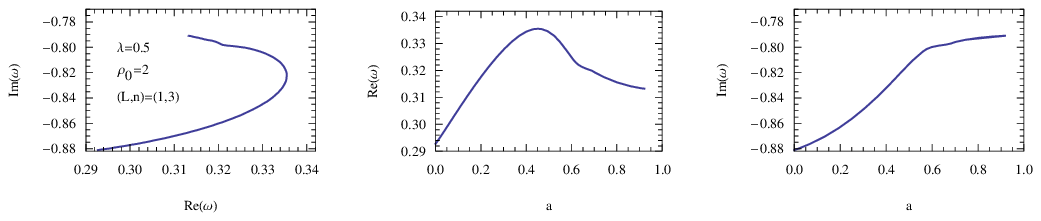}
\includegraphics[width=13cm]{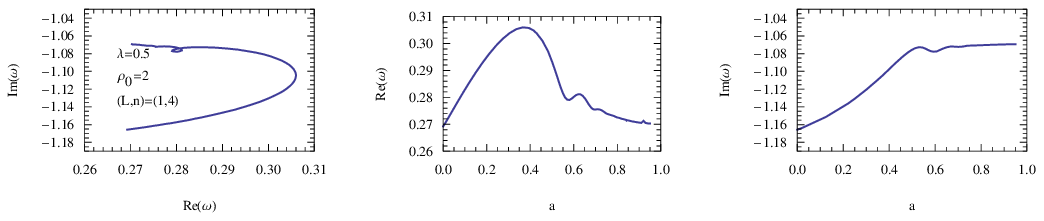}
\includegraphics[width=13cm]{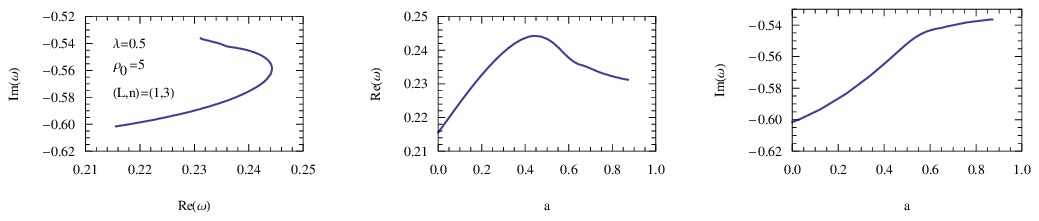}
\includegraphics[width=13cm]{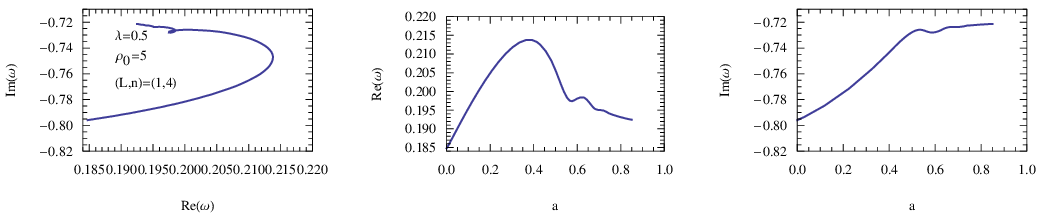}
\caption{\label{g132-142-135-145-0.5}the left four panels are
trajectories in the complex $\omega$ plane of scalar QNMs around
squashed KK black holes, for $n=3$, $n=4$, with $L=1$, $\lambda
=0.5$, $\rho_0=2$ and $5$. The others are real parts ($Re(\omega)$)
and imaginary parts $(Im(\omega)$) versus parameter $a$.}
\end{figure}

\begin{figure}
\includegraphics[width=13cm]{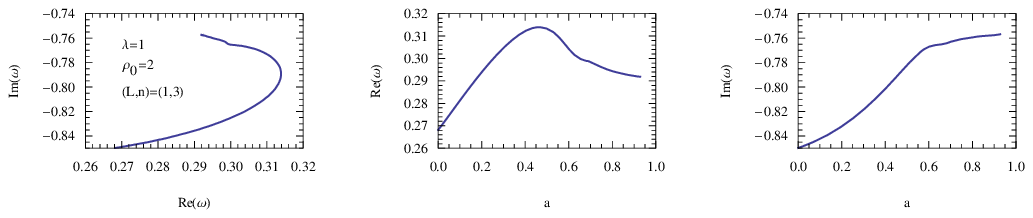}
\includegraphics[width=13cm]{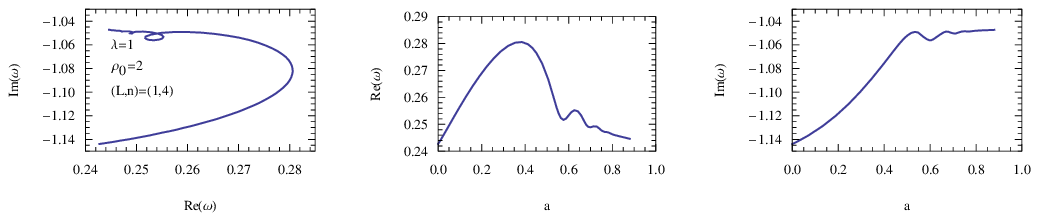}
\includegraphics[width=13cm]{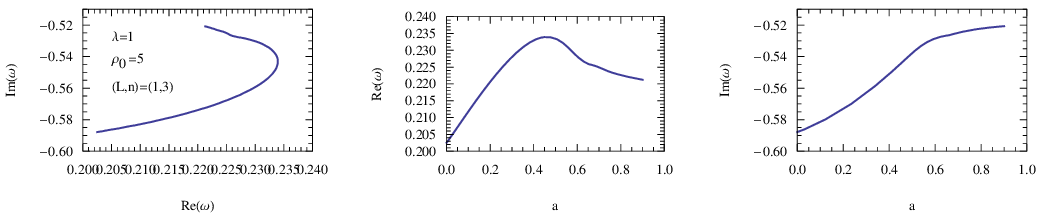}
\includegraphics[width=13cm]{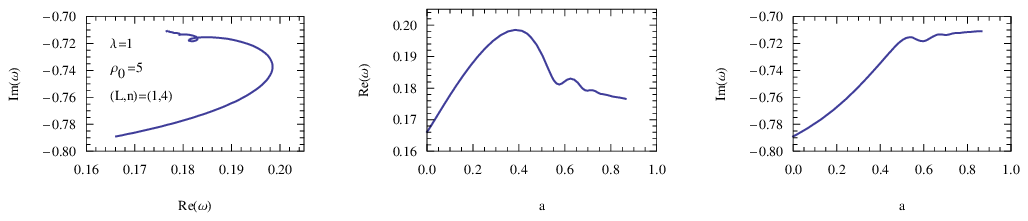}
\caption{\label{g132-142-135-145-1}the left four panels are
trajectories in the complex $\omega$ plane of scalar QNMs around
squashed KK black holes, for $n=3$, $n=4$, with $L=1$, $\lambda =1$,
$\rho_0=2$ and $5$. The others are real parts ($Re(\omega)$) and
imaginary parts $(Im(\omega)$) versus parameter $a$.}
\end{figure}

In the following we report our results for taking $\lambda\neq 0$.
Numerical calculation for the case $\lambda\neq 0 $ is much more
time consuming than the case with $\lambda=0$.  In
Fig.~(\ref{g132-142-135-145-0.5}), we show results for L=1, n=3, 4
with $\rho_0=2,~ 5$, respectively by choosing $\lambda=0.5$. In
Fig.~(\ref{g132-142-135-145-1}) we show results for the same choice
of L, n but with $\lambda=1$. From these figures, it is obvious that
the spiral-like shape still appears at the critical overtone number
$n_c$,  namely $n_c=4$ for $L=1$ with different $\rho_0$ and nonzero
$\lambda$ in the scalar perturbation. The critical moment to exhibit
the spiral behavior in QNM again agrees well with the Davies¡¯
thermodynamical point which is shown in Table \ref{error3}.

\begin{table}[!h]
\tabcolsep 0pt \caption{Numerical data compared with H. Ishihara's
work , with $\rho_0=0.5$, $\rho_+=1$.} \vspace*{-12pt}
\begin{center}
\def\temptablewidth{0.7\textwidth}
{\rule{\temptablewidth}{1pt}}
\begin{tabular*}{\temptablewidth}{@{\extracolsep{\fill}}ccccccc}
L & $\lambda$ &WKB(6th order) &Frobrnius &Leaver  \\
\hline
10 & 0     & 3.5194-0.15956i   & 3.510564-0.159674i & 3.5105636-0.1596744i   \\
10 & 1/2   & 3.5336-0.15821i   & 3.529225-0.158317i & 3.5292246-0.1583174i   \\
10 & 1     & 3.5898-0.15413i   & 3.585417-0.154231i & 3.5854166-0.1542309i    \\
10 & 3/2   & 3.6842-0.14729i   & 3.679778-0.147368i & 3.6797783-0.1473680i  \\
10 & 2     & 3.8178-0.13759i   & 3.813421-0.137647i & 3.8134213-0.1376469i\\
10 & 5/2   & 3.9924-0.12493i   & 3.988012-0.124944i & 3.9880116-0.1249440i  \\
10 & 3     & 4.2103-0.10912i   & 4.205922-0.109076i & 4.2059218-0.1090755i   \\
10 & 7/2   & 4.4747-0.08988i   & 4.470523-0.089745i
&4.4705233-0.0897455i
\end{tabular*}
{\rule{\temptablewidth}{1pt}}\label{compare}
\end{center}
\end{table}

\begin{table}[!h]
\tabcolsep 5mm \caption{The comparison of values of critical points
for QNMs ($a_{Q}$) and the singular point of the heat capacity
($a_{D}$), when $\lambda=0$. }
\begin{center}
\begin{tabular}{r@{.}lr@{.}lr@{.}l}
\hline
\multicolumn{2}{c}{$(\lambda,L,~n,~\rho_0)$}&\multicolumn{2}{c}{$\triangle
a=|a_{Q}-a_{D}|$}&\multicolumn{2}{c}{${\triangle a\over a_{Q}}$}
\\ \hline
(0,~0,~1,~0&)      &0&022       &4&21 $\%$  \\
(0,~0,~1,~1&)      &0&016       &3&85  $\%$   \\
(0,~0,~1,~2&)      &0&018       &3&85  $\%$     \\
(0,~1,~4,~0&)      &0&007      &0&953   $\%$ \\
(0,~1,~4,~1&)      &0&0082      &2&091  $\%$ \\
(0,~1,~4,~2&)      &0&0041      &1&105   $\%$ \\
\hline
\end{tabular}\label{error}
\end{center}
\end{table}

\begin{table}[!h]
\tabcolsep 5mm \caption{The comparison of values of critical points
for QNMs ($a_{Q}$) and the singular point of the heat capacity
($a_{D}$), when $\lambda=0.5$ and $\lambda=1$. }
\begin{center}
\begin{tabular}{r@{.}lr@{.}lr@{.}l}
\hline
\multicolumn{2}{c}{$(\lambda,~L,~n,~\rho_0)$}&\multicolumn{2}{c}{$\triangle
a=|a_{Q}-a_{D}|$}&\multicolumn{2}{c}{${\triangle a\over a_{Q}}$}
\\ \hline
(0.5,~1,~4,~2&)      &0&0035       &0&942 $\%$  \\
(0.5,~1,~4,~5&)      &0&0237       &6&29  $\%$   \\
(~1~,~1,~4,~2&)      &0&0001       &0&027 $\%$  \\
(~1~,~1,~4,~5&)      &0&0330        &8&56  $\%$   \\
\hline
\end{tabular}\label{error3}
\end{center}
\end{table}

In summary we have studied the QNM of scalar perturbation in the
background of the charged KK black hole with squashed horizons. We
observed  that over some critical value of the black hole
parameter, both the real part and the imaginary part of the QNMs
will experience oscillations and the complex $\omega$ plan will
exhibit the spiral-like shape. Interestingly this critical value
agrees well to the Davies' point on the thermal stability of black
holes obtained from the singular position of the heat capacity.
Our limiting case returns to the RN black hole, where the relation
was observed in \cite{17}. The correspondence of the critical
point observed in QNM to the position of infinite discontinuities
of the heat capacities indicates that QNM may shed the light on
the turning point of the thermal stability. It is of interest to
generalize this discussion to other backgrounds and for different
fields' perturbations.

For a black brane solution it was conjectured that black holes which
lack local thermodynamical stability often also lack stability
against small perturbations\cite{23}. This conjecture might only
hold for black holes with translation symmetry, such as black
string. In our background spacetime we observed that  on both sides
of the turning point of the thermodynamical stability, the imaginary
frequencies of the QNM are negative, which tells us that the scalar
perturbation is always stable even at the critical point. Recently
the metric perturbations were discussed \cite{xx,Ishihara} and it
was indicated that dynamically the squashed KK black hole is stable,
this is consistent with our result. This result shows that in the
background we are studying it has dynamical stability even when
thermodynamical instability appears. This is quite interesting,
which suggests that the relation between the dynamical and
thermodynamical stabilities of black holes is non-trivial and
further investigations are called for.

\begin{acknowledgments}

We thanks J.L. Jing and Q.Y. Pan for helpful discussions. This
work was partially supported by NNSF of China, Shanghai Education
Commission and Shanghai Science and Technology Commission. S. B.
Chen's work was partially supported by the China Postdoctoral
Science Foundation under Grant No. 20070410685, the Scientific
Research Fund of Hunan Provincial Education Department Grant No.
07B043 and the National Basic Research Program of China under
Grant No. 2003CB716300. B. Wang would like to acknowledge the
associate programme in ICTP where the work was completed.
\end{acknowledgments}

\end{document}